\documentclass[oldversion]{aa}

\usepackage{graphicx}
\usepackage{epstopdf}
\usepackage{natbib}
\bibpunct{(}{)}{;}{a}{}{,}

\begin{document}

\title{The influence of AGN nuclear parameters on the FRI/FRII dichotomy}

\author{M. Wold \inst{1,2,3} \and M. Lacy \inst{2} \and L. Armus \inst{2}}

\offprints{M. Wold}

\institute{European Southern Observatory, Karl-Schwarzschild str.\ 2,
           85748 Garching bei M{\"u}nchen, Germany
	   \and 
           Spitzer Science Center, California Institute of Technology, 
           MS 220-6, Pasadena, 
	   CA 91125, U.S.A.
           \and
           Present address: Institute of Theoretical Astrophysics, P.O.\ Box
           1029 Blindern, N-0315 Oslo, Norway}

\date{Received ; Accepted}

\abstract{We have investigated the influence of nuclear parameters
  such as black hole mass and photoionizing luminosity on the FRI/FRII
  transition in a sample of nearby ($z<0.2$) radio galaxies from the
  3CR catalogue.  The sample was observed with medium-resolution,
  optical spectroscopy and contains some galaxies with unpublished
  velocity dispersion measurements and emission-line fluxes.  The
  measured velocity dispersions for the sample lie in the range
  130--340 km\,s$^{-1}$ with a mean of 216 km\,s$^{-1}$.  Using the
  M-$\sigma$ relation, we convert to black hole mass and find that the
  black hole mass distribution is identical for FRI and FRII galaxies,
  with a mean of $\approx2.5\times10^{8}$ M$_{\odot}$. We determine
  narrow emission-line luminosities from [\ion{O}{ii}] and
  [\ion{O}{iii}] in our spectra, as well as from the literature, and
  convert them to photoionizing luminosities under the assumption that
  the gas is ionized by the nuclear UV continuum. Most of the galaxies
  with FRI morphology and/or low-excitation emission-line spectra have
  progressively lower black hole masses at lower photoionizing (and
  jet) luminosities. This agrees with the well-known Ledlow-Owen
  relation which states that the radio luminosity at the FRI/FRII
  transition depends on the optical luminosity of the host, $L_{\rm
    radio}\propto L_{\rm optical}^{1.8}$, because these two
  luminosities relate to AGN nuclear parameters.  When recasting the
  Ledlow-Owen relation into black hole mass versus photoionizing
  luminosity and jet luminosity, we find that the recasted relation
  describes the sample quite well.  Furthermore, the FRI/FRII
  transition is seen to occur at approximately an order of magnitude
  lower luminosity relative to the Eddington luminosity than the
  soft-to-hard transition in X-ray binaries. This difference is
  consistent with the Ledlow-Owen relation, which predicts a weak
  black hole mass dependence in the transition luminosity in Eddington
  units.  We conclude that the FRI/FRII dichotomy is caused by a
  combination of external and nuclear factors, with the latter
  dominating.}

\maketitle

\section{Introduction}

Radio galaxies are usually classifed as FRI or FRII sources depending
on their radio morphology. FRIs have smooth jets emanating from the
nucleus and lobes where the surface brightness decreases toward the
edges. The more powerful FRIIs have faint jets and edge-brightened
lobes.  The classification scheme was invented by \citet{fr74}, who
discovered that the FRI/FRII transition occurs at a radio luminosity
of $P_{\rm 178 MHz} \approx 10^{25}$ W\,Hz$^{-1}$\,sr$^{-1}$, with
almost all sources below the transition luminosity being FRIs.

Over the years, much work has been put into understanding the
remarkable FRI/FRII transition. Models fall into two different groups,
those explaining the morphological differences as arising because of
different physical conditions in the environment in which the radio
source propagates, and those seeking to explain the dichotomy as
caused by fundamental AGN parameters or the jets (see \citet{gkw00}
for a summary).  The deceleration models, belonging to the former
group, have been successful in reproducing the appearance of both FRI
and FRII jets. In this scenario, the jets are thought to start out
supersonically and slow down to a tran/subsonic flow because of
entrainment of plasma in the host galaxy
\citep{deyoung93,laing94a,laing96,bicknell95,ka97}.  The other group
of models explains the differences as arising in more fundamental
parameters like black hole (BH) spin, accretion mode, or jet
composition
\citep{rees82,baum95,reynolds96,meier99,meier01,gc01,marchesini04}.

An interesting discovery was made by \cite{lo96} (see also
\citealt{ol89,ow91,ol94}) who found that the FRI/FRII transition radio
luminosity is an increasing function of the optical luminosity of the
host galaxy, $L_{\rm rad} \propto L_{\rm opt}^{2}$. This means that
the more optically luminous a galaxy is, the more powerful its radio
source must be in order to produce FRII morphology.  The Ledlow-Owen
relation has been an important observation in the effort to explain
radio galaxy dichotomy.  However, both groups of models are able to
explain, or reproduce, the Ledlow-Owen relation
\citep{bicknell95,meier99,gkw01}.

From the correlation between bulge luminosity and BH mass
\citep{kr95,magorrian98}, we expect the host optical luminosity to
scale with BH mass.  Furthermore, because the radio-optical
correlation (optical here means nuclear optical luminosity) for radio
galaxies \citep{saunders89,rs91,willott99,grimes04} relates radio
luminosity to narrow-line luminosity, the radio luminosity can be
expressed as a function of nuclear photoionizing luminosity. The
Ledlow-Owen relation can thus be cast into the variables BH mass and
nuclear photoionizing luminosity. \cite{gc01} use the above line of
reasoning to argue that the Ledlow-Owen relation reflects a change in
the accretion mode between the two classes because the transition occurs
at a fixed ratio between BH mass and photoionizing/accretion
luminosity.  Recent work by \citet{marchesini04} also argues for a
change in the accretion mode.  They study a sample of radio-loud
quasars and radio galaxies, finding that whereas quasars and radio
galaxies have a similar distribution in BH mass, the distribution of
accretion rates is bimodal with FRIs and weak-lined FRIIs on one side,
and FRIIs and radio-loud quasars on the other.

In the above-mentioned work, optical luminosity is used as a measure
of photoionizing luminosity. It is however still a matter of
controversy whether this is applicable to FRI galaxies because it is
uncertain whether they have dusty torii that can obscure the nucleus,
leading to an underestimate of their true optical luminosity. One way to
avoid the ambiguity in the interpretation of optical (and X-ray)
nuclear luminosity is to use narrow emission lines as these
are believed to arise from the narrow-line region outside the torus
and be photoionized by the UV continuum from the central AGN.

In this paper we revisit the Ledlow-Owen relation for a sample of
nearby FRI and FRII galaxies by utilizing stellar velocity dispersions
to estimate BH masses and narrow emission-line luminosities to
estimate photoionizing luminosities. We assume a cosmology with
$H_{0}=70$ km\,s$^{-1}$\,Mpc$^{-1}$, $\Omega_{m}=0.3$ and
$\Omega_{\Lambda}=0.7$ unless otherwise noted.  The abbreviations
[\ion{O}{ii}] and [\ion{O}{iii}] are used for [\ion{O}{ii}]3727 {\AA}
and [\ion{O}{iii}]5007 throughout the paper.


\section{The sample}

The sample consists of sources drawn from the complete 3CRR catalogue
of \cite{lrl83}. The selection criterion was $P_{178} \leq 10^{26.5}$
W\,Hz$^{-1}$\,sr$^{-1}$, hence covering the FRI/FRII transition
regime. Further constraints on RA and DEC gave a sample of 21 targets,
ten FRIs and eleven FRIIs. Because of the redshift-luminosity
correlation in the 3CRR sample, all targets lie at $z < 0.2$.  The
properties of the selected sources are listed in Table~\ref{table:t1}.
Radio power at rest-frame 178 MHz was calculated from the 178 MHz flux
and the spectral index as listed in the 3CRR Atlas web pages by Leahy,
Bridle \& Strom\footnote{http://www.jb.man.ac.uk/atlas}. Radio
morphology class was taken from the same web pages.  In addition to
morphology, we also list emission-line class as either high-excitation
('HEG') or low-excitation ('LEG') emission line galaxy. FRIIs are
classified as HEGs if [\ion{O}{iii}] is strong compared to
[\ion{O}{ii}], and as LEGs if [\ion{O}{iii}] is weak or absent
\citep{laing94b}. FRIs have low-excitation spectra, hence all the FRIs
are listed as LEGs in Table~\ref{table:t1}.

Emission-line classes for all objects were found in the literature
(see table note) except for 3C~442A, which we classify as a LEG
because [\ion{O}{iii}] is weaker than [\ion{O}{ii}] in our spectrum.
This galaxy is not a classical FRII with well-defined hot spots (one
of the lobes has FRI morphology), and the host galaxy is part of an
ongoing merger \citep{co91}. The other source with non-classical radio
morphology in the sample is 3C~433, highly asymmetric and probably
interacting with close neighbours. The radio morphology is classified
as FRI by Leahy, Bridle \& Strom, but we have chosen to define it as
an FRII because its optical spectrum is characteristic of an HEG and
there is evidence that it harbours an obscured quasar nucleus
\citep{fabbiano84,lilly85,yl89}.

\begin{figure}
\begin{center}
\includegraphics{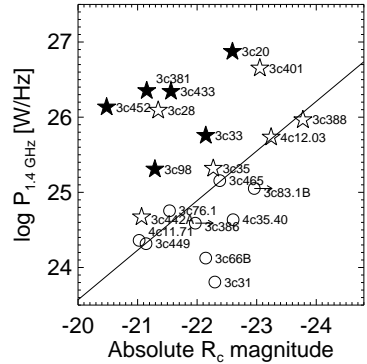}
\caption{ Open circles correspond to FRIs and stars to FRIIs,
  with filled symbols for high-excitation emission line galaxies.  The
  solid line marks the Ledlow-Owen relation.}
\label{figure:lo}
\end{center}
\end{figure}

In order to compare the sources with the Ledlow-Owen relation, we plot
the sample in the radio-optical plane in Fig.~\ref{figure:lo}. The
apparent rest-frame $R_{c}$ magnitudes were converted to absolute
magnitudes, and radio luminosity at 1.4 GHz calculated using the
observed flux at 1.4 GHz and spectral indices as listed in
Table~\ref{table:t1}.  For calculating luminosities, we used the same
cosmology as \citet{lo96}. The figure shows that the galaxies fall in
the expected regions of the diagram, with FRI sources below and FRII
sources above the Ledlow-Owen relation, but there are also some
borderline cases. All galaxies classified as HEGs are well above the
relation, consistent with the unified schemes \citep{jw99,ccc02}.

\begin{table*}
  \caption{The sample. 
    $P_{\rm 178}$ in column (3) is total radio power (W\,Hz$^{-1}$sr\,$^{-1}$) 
    at rest-frame 178 MHz. Column (4) lists spectral index between 178 and 
    750 MHz, and columns (5) and
    (6) the radio morphology and emission-line type. 
    Apparent magnitude in rest-frame $R_{C}$-filter and corresponding
    references are listed in column (7) and (8). Columns (9) and (10) show
    velocity dispersion measurements from the literature (km\,s$^{-1}$) 
    and corresponding references. The last three columns list emission line 
    fluxes in units of 
    10$^{-15}$ erg\,s$^{-1}$\,cm$^{-2}$, both from the literature
    and as determined from our wide slit spectrophotometry, and references.}
\centering
\begin{tabular}{lllllllllllll}
\hline\hline
Radio ID  & Redshift & $\log P_{178}$& $\alpha_{178}^{750}$ 
    &\multicolumn{2}{c}{ } & $m_{R_{C}}$ &  Ref & $\sigma_{\rm lit}$ & Ref & $S_{\rm [OII]}$ & $S_{\rm [OIII]}$ & Ref \\
   (1)     &  (2)      &     (3)      &  (4) & (5) &  (6)  &  (7) & (8) & (9) & (10) & (11) & (12) & (13) \\
\hline
3C 20    &   0.174  & 26.49 & 0.66 & FRII & HEG & 16.80 & 1 & & & 0.634 & 1.915 & W \\
3C 28    &  0.1971 & 26.22 & 1.06 & FRII & LEG & 18.34 &  2 & & & 5.956 & ... & W \\
3C 31    &  0.0167 & 23.99 & 0.57 & FRI  & LEG & 11.92 & 2 & 249 & SHI90 & ... & 10.349 & G \\
3C 33    &  0.0595 & 25.60 & 0.76 & FRII & HEG & 14.80 & 3 & & & 33.017 & 147.48 & W \\
3C 35    &  0.0673 & 24.99 & 0.77 & FRII & LEG & 14.94 & 1 & & & 1.0$\pm$0.2 & $<0.34$ & A \\
3C 66B   &  0.0215 & 24.35 & 0.50 & FRI & LEG & 12.55 & 2 & & & 6.782 & 4.801 & W \\
3C 76.1  &  0.0324 & 24.41 & 0.77 & FRI & LEG & 14.06 & 2 & 246 & SHI90 & $<3.0$ & $<2.0$ & A \\
3C 83.1B &  0.0255 & 24.53 & 0.62 & FRI & LEG & $<12.11$  & 4 & & & & & \\
3C 98    &  0.0306 & 24.95 & 0.78 & FRII & HEG & 14.18 & 3 & 175 & SHI90 & 7.936 & 39.775 & W \\
3C 381   &  0.1605 & 26.01 & 0.81 & FRII & HEG & 18.05 & 5 & & & 1.177 & 44.31 & W \\
3C 386   &  0.0170 & 24.13 & 0.59 & FRI & LEG & $<$12.21 & 4 & & & & & \\
3C 388   &  0.0908 & 25.64 & 0.70 & FRII & LEG & 14.12 & 3 & 365 & H85 & 1.618 & 1.901 & W \\ 
3C 401   & 0.201  & 26.32 & 0.71 & FRII & LEG & 16.68 & 6 & & & ... & 1.192 & W \\
3C 433   & 0.1016 & 26.10 & 0.75 & FRII & HEG & 16.59 & 5 & & & ... & 2.656 & W \\
3C 442A  & 0.0263 & 24.35 & 0.96 & FRII & LEG & 14.07 & 2 & 197 & SHI90 & 8.0$\pm$0.8 & 4.7$\pm$1.0 & A \\
3C 449   &  0.0171 & 23.82 & 0.58 & FRI  & LEG & 13.05 & 2 & 222 & SHI90 & $<3.0$ & $<2.0$ & A \\
3C 452   & 0.0811 & 25.88 & 0.78 & FRII & HEG & 17.16 & 7 & & & 17.149 & ... & W\\
3C 465   & 0.0305 & 24.87 & 0.75 & FRI  &  LEG & 13.14 & 2 & 341 & FIF95 & 2.0$\pm$0.4 & $<4.0$ & A \\
4C 11.71 & 0.0262  & 24.12 & 0.75 & FRI  & LEG & 14.00 & 7 & 249 & EFAR99 & $<5.0$ & $<3.0$ & A \\
4C 12.03 & 0.157  & 25.77 & 0.87 & FRII & LEG & 15.91 & 1 & & & ... & 1.394 & W \\
4C 35.40 &  0.0301 & 24.29 & 0.76 & FRI  & LEG & 12.83 &  8$^{\mathrm{ }}$ & 204 & EFAR99 & 1.0$\pm$0.2 & $<1.3$ & A \\
\hline
\end{tabular}

\begin{list}{}{}
\item[$^{\mathrm{ }}$] References for $m_{R_{C}}$: (1) Converted from
  $K_{\rm UKIRT}$ \citep{ll84} assuming $R_{C}-K=2.5$
  \citep{dunlop03}, (2) \cite{cr04}, (3) \cite{ol89}, (4)
  \cite{martel99} (upper limit given because of contamination by
  foreground star), (5) converted from $V$ \citep{sh89} assuming
  $V-R_{C}=0.61$ \citep{fukugita95}, (6) converted from $K$
  \citep{lebofsky81} assuming $R_{C}-K$=2.5 \citep{dunlop03}, (7)
  converted from $V$ \citep{sandage73} assuming $V-R_{C}=0.61$
  \citep{fukugita95}, (8) converted from 2MASS $K_{s}$ assuming
  $R-K_{s}=2.5$ \citep{dunlop03}. Emission-line classifications are
  from \cite{jr97}, except for 3C~28 and 4C~12.03 which are from
  Willott's compilation ({\em
    http://www-astro.physics.ox.ac.uk/\~\,cjw/3crr/3crr.html}), and
  3C~442A which we classify as LEG based on our Palomar spectrum.  The
  1.4 GHz fluxes were taken from NVSS by \cite{condon98} and from the
  surveys of \cite{wb92} and \cite{lp80}.  References for velocity
  dispersions: SHI90: \citet{smith90}, H85: \citet{heckman85}, FIF95:
  \citet{fif95} and EFAR99: \citet{efar99}.  References for
  emission-line fluxes: A: from our data, W and G: from Willott's and
  Grimes' compilations, respectively ({\em
    http://www-astro.physics.ox.ac.uk/\~\,cjw/3crr/3crr.html} and {\em
    http://www-astro.physics.ox.ac.uk/\~\,sr/grimes.html}).
\end{list}

\label{table:t1}
\end{table*} 


\section{Observations and data reduction}

Medium resolution ($R\sim3000$--4000) optical spectra were obtained
with the Double Spectrograph \citep{og82} on the 5m Hale Telescope at
Palomar Observatory. The spectrograph was equipped with two
1k$\times$1k CCDs and the D48 dichroic, dividing the red and blue side
at approximately 4800 {\AA}. We utilized the 1200 l/mm grating blazed
to 7100 {\AA} on the red CCD and to 4700 {\AA} on the blue CCD. Two
different grating angles were used depending on the galaxy redshift
such that the \ion{Mg}{I}{\em b} $\lambda$5175 {\AA} absorption line
complex fell on the center of the red CCD, and the Ca H \& K
$\lambda\lambda$3934,3969 {\AA} doublet at the center of the blue CCD.
The red and the blue CCDs had pixel scales of 0\farcs468 and
0\farcs624, resulting in dispersions of 0.624 and 0.864 {\AA}/pix,
respectively.

Absorption line spectra were obtained with a 1\arcsec\, wide slit,
giving spectral resolutions of 1.33 and 1.39 {\AA}, corresponding to
velocity resolutions of $\approx70$ and $\approx100$ km\,s$^{-1}$ in
the red and the blue, respectively. In order to measure total narrow
emission line fluxes, we also took spectra with a wide slit (4\arcsec)
of part of the sample. Slit lengths were 128\arcsec.  Spectra were
taken by centering the slit on the optical nucleus of the galaxy and
aligning it with the major axis in order to maximize the amount of
light entering the slit. Each target was also observed as close to the
parallactic angle as possible in order to minimize losses due to
differential atmospheric refraction. Details about the observations
are given in Table~\ref{table:t3}.

Wavelength calibration was accomplished by taking arc lamp spectra at
every telescope pointing and standard stars were observed at the
beginning and end of each night. There were some clouds at the
beginning of the first night, but for the rest of the run the weather
was clear and photometric, and the seeing stable at $\approx1$\arcsec.
The photometric zero point varied 0.1--0.2 mag between the beginning
and end of the second and third night. For the first night, which
started out as non-photometric, the difference was $\approx0.7$ mag.
For spectrophotometry with the 4\arcsec\, slit, we therefore waited
until the middle of the first night when the conditions were more
favourable.  As templates for the velocity dispersion fitting we
obtained spectra of one sub giant G0 star (HD188121) and one giant K1
star (HD28191).  In order to avoid seeing-limited spectral resolution
with the 1\arcsec\, slit, the template star was moved across the slit
during integration.

\begin{table}
\caption{Table of observations. For each galaxy, the slit position angle 
and exposure time is shown.}
\centering
\begin{tabular}{lrlrrl}
\hline\hline
   & \multicolumn{2}{c}{1\arcsec\, slit} & \multicolumn{2}{c} 
{4\arcsec\, slit} &  \\
Galaxy    & PA   & $T_{\rm exp}$ & PA  & $T_{\rm exp}$ & Date \\
          & (deg) & (sec) & (deg) & (sec)    &  \\
\hline		
3C 20     & 125   & 3600     & ...     & ... & 11Sep02 \\
3C 28     & 59    & 2400     & ...     & ... & 11Sep02 \\
3C 31     & 144   & 2400     & ...     & ... & 10Sep02 \\
3C 33     & 146   & 2400     & ...     & ... & 10Sep02 \\
3C 35     & 111   & 2400     & 23          & 900     & 10Sep02 \\
3C 66B    & 135   & 1200     & ...     & ... & 10Sep02 \\
3C 76.1   & 135   & 2400     & 130         & 900     & 12Sep02 \\
3C 83.1B$^{\mathrm{a}}$  & 166 & 1800 & 96 & 2400 & 12Sep02 \\ 
3C 98     & 150 & 2400     & ...    & ...       & 12Sep02 \\
3C 388    & 71	  & 2400     & 71      & 900     & 11Sep02 \\
3C 381    & 154   & 3600     & ...    & ...      & 11Sep02 \\
3C 386$^{\mathrm{b}}$

  & 12    & 2400     & 12         & 1800         & 10Sep02 \\
3C 401    & 0     & 3600     & 0    & 1200    & 11Sep02 \\
3C 433    & 66    & 2400     & 	...    &  ...       & 11Sep02 \\ 
3C 442A   & 131   & 2400     & 57   & 900     & 12Sep02 \\
3C 449    & 1     & 2400     & 8    & 900     & 10Sep02 \\
3C 452    & 0     & 2400     & ...   & ... & 12Sep02 \\
3C 465    &  30   & 1800     & 121  & 900     & 12Sep02 \\
4C 11.71  &  326  & 2400     & 53   & 900     & 10Sep02 \\
4C 12.03  &  17   & 2400     & 17  & 900     & 11Sep02 \\
4C 35.40  & 84    & 2400     & 133  & 1200    & 12Sep02 \\
\hline
\end{tabular}
\label{table:t3}

\begin{list}{}{}
\item[$^{\mathrm{a}}$] Confused with a star 
3\arcsec\, away from the nucleus \citep{poulain92,dekoff00}.
\item[$^{\mathrm{b}}$] A star is superposed on the galaxy nucleus
\end{list}

\end{table}

The spectra were reduced and extracted in a standard manner using {\sc
  iraf}\footnote{{\sc iraf} is distributed by the National Optical
  Astronomy Observatories, which are operated by the Association of
  Universities for Research in Astronomy, Inc., under cooperative
  agreement with the National Science Foundation.} tasks in the {\em
  twodspec} and {\em onedspec} packages. Night sky lines from
\ion{O}{I} and from mercury in street-lamps were used to check the
wavelength calibration, and an agreement of typically $\la 0.5$ pixels
was found, corresponding to $\la 18$ and $\la 32$ km\,s$^{-1}$ in the
red and blue, respectively.

Two of the spectra, 3C~83.1B and 3C~386, are dominated by light from
foreground stars. For this reason we do not consider these two
galaxies in our analysis. In the case of 3C~83.1B, the star lies
3\arcsec\, to the east of the galaxy nucleus
\citep{poulain92,dekoff00}, and for 3C~386 the star is superposed on
the optical galaxy core \citep{ccc02}.


\section{Template fitting procedure}

We fit stellar templates to the galaxy spectra in order to measure
velocity dispersions using the direct fitting method described by
\citet{barth02}.  Galaxy and stellar template spectra were extracted
using aperture diameters of $1\farcs 872$ in order to cover the same
physical scale in both spectral arms. A model spectrum, $M(x)$, where
$x=\ln \lambda$ is measured in the galaxy rest frame was evaluated as
\begin{equation}
M(x) = \left\{ \left[T(x) \otimes G(x)\right] + C(x) \right\} \times P(x),
\end{equation}
\noindent
where $[T(x) \otimes G(x)]$ denotes the template spectrum convolved
with a Gaussian of width $\sigma_{*}$, $C(x)$ is the continuum and
$P(x)$ a polynomial.  The best-fit model was found by minimizing
$\chi^{2}$ using a downhill simplex method implemented in the {\em
  amoeba} algorithm \citep{press88}.
 
We experimented with different functional forms of $C(x)$ and $P(x)$,
starting with a straight line for the continuum, and a Legendre
polynomial for $P(x)$.  However, since the change in $x$ is small over
the fitting range, coefficient degeneracies were easily introduced,
hence we rescaled the fitting range to a new variable $\theta
\in[0,\pi]$ and used Fourier-type expansions instead.  The continuum
is then expressed as
\begin{equation}
C(\theta)=c_{0}+c_{1}\sin\theta,
\end{equation}
\noindent
and the polynomial as 
\begin{equation}
P(\theta)  =  p_{0} + p_{1}\cos\theta + p_{2}\cos2\theta + 
	      p_{3}\cos3\theta + p_{4}\cos4\theta.
\end{equation} 
The sine and cosine terms are orthogonal, thereby breaking part of the
degeneracy.

\begin{figure*}
\centering
\includegraphics[width=17cm]{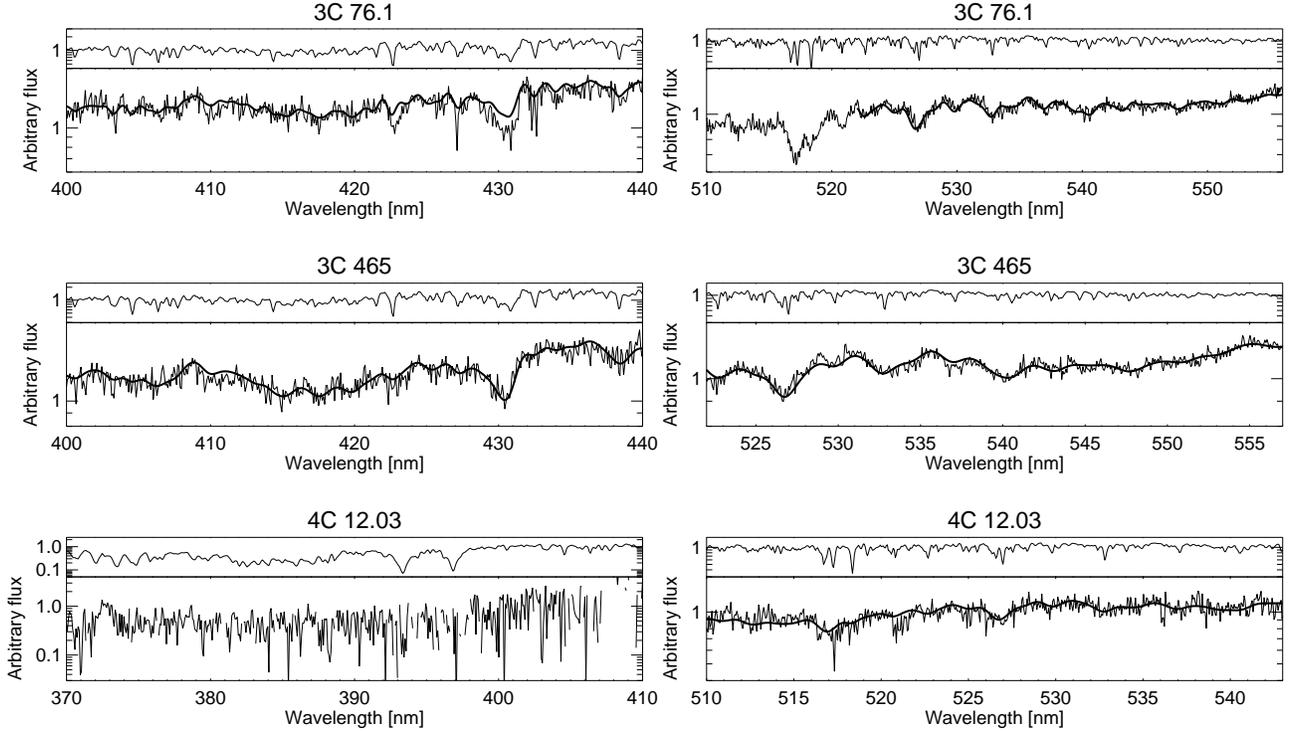}
\caption{Examples of radio galaxy spectra with best-fit models
  overplotted as solid lines where an acceptable fit could be
  obtained. The K1 star used as a template for the fitting is shown at
  the top of each panel.}
\label{fig:results}
\end{figure*}

Fits to three different galaxies along with the stellar template used
to construct the model are shown in Fig.~\ref{fig:results}. Generally,
we found that the regions redward of Ca H\&K and the \ion{Mg}{I}{\em
  b} absorption line complex are best suited for fitting velocity
dispersions.  This has also been experienced in other works, e.g.\
\cite{barth02} and \cite{gh06}.  We therefore tried to avoid the
\ion{Mg}{I}{\em b} region, except for in some cases where the Fe
blends redward of the region were not sufficiently covered.  As Barth
et al., we also find excess emission at $\approx 5200$ {\AA} in
several cases (caused by [\ion{N}{I}]5199 {\AA} emission in the host),
complicating the fitting.  For the red arm spectra, we therefore
started several of the fits at $\approx 5220$ {\AA}.  The results from
the fitting and the fitting regions used are listed in
Table~\ref{table:t4}.

\begin{figure}
\begin{center}
\includegraphics{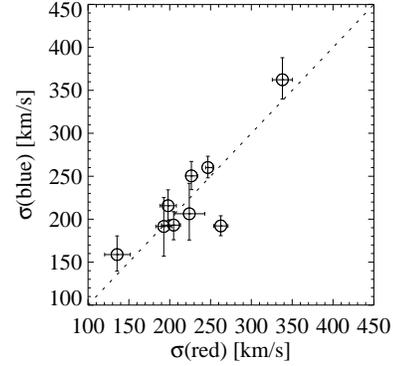}
\caption{Comparison of red and blue velocity dispersions. Error bars
  represent 95 \% confidence intervals.}
\label{fig:vdisp_comparison}
\end{center}
\end{figure}

The red arm spectra were easier to fit than those from
the blue arm, probably a combination of the red spectra having
better signal-to-noise and containing less spectral features to fit.
Also, because the galaxies contain spectral features that match better
with K stars than with G stars in the chosen fitting regions, the K1
template was found to produce better fits with less scatter than the
G0 template.  As there is also less contamination by the AGN continuum
in the red part, we chose to use the red arm spectra with the K1
template. In Fig.~\ref{fig:vdisp_comparison} we show a comparison
between measurements from the red and the blue arm in cases where
spectra from both could be fitted. The errors are seen to be larger
for the blue fits, and the two agree to within typically ten per cent,
except for two cases, 3C~449 and 3C~98 where there are differences at
the 20--30\% level.

\begin{table*}
  \caption{Results from the fitting. The fitting regions and signal-to-noise 
    per pixel in the continuum is listed. Best-fit  velocity dispersions 
    normalized to the same physical aperture
    ($r_{\rm ap} = 0.595h^{-1}$ kpc) and their confidence intervals are shown. 
    'AGN cont.' indicates that there is too much AGN contamination 
    in the spectrum to obtain a reliable fit.}
\centering
\label{table:t4}
\begin{tabular}{llrlll|lrlll}
\hline\hline
          & Fitting region& S/N & $\sigma_{*}$ & 68 \% CI & 95 \% CI & Fitting region & S/N & $\sigma_{*}$ & 68 \% CI & 95 \% CI \\
          &    {\AA} &    &     km\,s$^{-1}$ &   km\,s$^{-1}$  &  km\,s$^{-1}$ & {\AA} &  & km\,s$^{-1}$ & km\,s$^{-1}$ & km\,s$^{-1}$ \\
\hline		

3C 20     & AGN cont.     & 4  & ... & ... & ... 
          & too noisy     & 1  & ... & ... & ... \\
3C 28     & AGN cont.     & 6  & ... & ... & ...
          & too noisy     & 1  & ... & ... & ... \\
3C 31 & 5220--5650 & 25 & 247 & [246,251] & [244,252] & 4000--4500 & 7 & 260 & [255,266] & [248,273] \\
3C 33 & 5220--5420 & 11 & 188 & [178,199] & [168,210] & AGN cont.  & 6 & ... & ... & ... \\
3C 35 & 5090--5390 & 6 & 224 & [215,233] & [206,243] & 3800--4250 & 3 & 206 & [191,224] & [176,242] \\
3C 66B$^{\mathrm{a}}$ & poor fit  & 18 & ... & ... &  ... & 4000--4470 & 6 & 165 & [155,174] & [146,185] \\
3C 76.1 & 5220--5560 & 18 & 193 & [188,198] & [183,203] & 4000--4470 & 7 & 192 & [174,209] & [157,225] \\
3C 98 &  5220--5570 & 17 & 136 & [128,143] & [120,152] & 4000--4420 & 7 & 159 & [149,170] & [140,181] \\
3C 388 & 5150--5740 & 15 & 232 & [211,243] & [195,259] & too noisy & 2 & ... & ... & ... \\
3C 381 & 5100--5410 & 9 & 221 & [212,231] & [202,241] & too noisy & 1 & ... & ... & ... \\
3C 401 & 5100--5230 & 5 & 136 & [95 ,166] & [65 ,284] & too noisy & 1 & ... & ... & ... \\
3C 433 & 5220--5700 & 10 & 128 & [121,131] & [115,136] & too noisy & 1 & ... & ... & ... \\
3C 442A & 5100--5590 & 15 & 198 & [188,208] &  [193,203] & 4000-4450 & 6 & 216 & [207,225] & [198,234] \\
3C 449 & 5100--5650 & 17 &  263 & [258,267] & [254,271] & 3800--4500 & 6 & 192 & [187,198] & [181,204] \\
3C 452 & 5120--5310 & 7 &  289 & [284,296] & [274,307] & AGN cont. & 4 & ... & ... & ... \\
3C 465 & 5220--5570 & 19 &  338 & [332,344] & [326,350] &  4000--4400 & 7 & 362 & [351,375] & [340,388] \\
4C 11.71 & 5220--5600 & 8 &  227 & [224,232] & [220,234] & 4000--4450 & 6 & 250 & [242,259] & [234,267] \\
4C 12.03 & 5100--5420 & 8 &  245 & [231,257] & [219,271] & too noisy & 1 & ... & ... & ... \\  
4C 35.40 & 5220--5570 & 16 & 205 & [198,205] & [190,213] & 4000--4420 & 6 & 193 & [184,200] & [176,209] \\
                                               
\hline
\end{tabular}

\begin{list}{}{}
\item[$^{\mathrm{a}}$] Despite good signal-to-noise in the red part,
  for unknown reasons, an acceptable fit could not be obtained.
\end{list}
\end{table*}

\begin{figure}
\begin{center}
\includegraphics{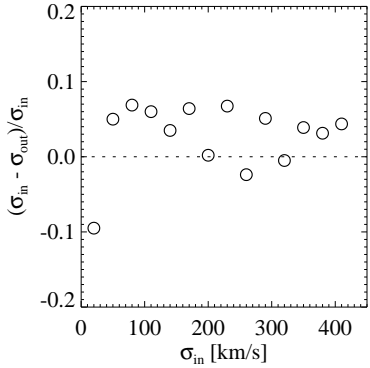}
\caption{Results from simulations with artificial galaxy spectra. The
  input velocity dispersions, $\sigma_{\rm in}$, are recovered
  ($\sigma_{\rm out}$) to within 10\%.}
\label{fig:simul}
\end{center}
\end{figure}

We tested the fitting routine by constructing artificial galaxy
spectra from the two stellar templates (50\% K1- and 50\% G0-star),
broadening them by a specified amount and applying the fitting
routine. Noise spectra were also added, broadened to the instrumental
resolution ($\approx 45$ km\,s$^{-1}$) and scaled in order to obtain a
signal-to-noise ratio of approximately 15.  The result from the
simulations is shown Fig.~\ref{fig:simul}, where it can be seen that
the velocity dispersions are generally recovered to within 10\% of the
input value. As expected, at dispersions lower than the instrumental
resolution there is larger scatter.

Nine of the radio galaxies in the sample have velocity dispersions
from the literature, as listed in column nine of 
Table~\ref{table:t1}.  A comparison with the literature measurements
is shown in Fig.~\ref{fig:vdisp_comparison}. There is agreement to
within less than 20\%, except for one galaxy, 3C~388, where our
estimated velocity dispersion is 36\% lower than the literature value.
This galaxy is one of the objects in our sample with noisier spectra,
hence it has larger error bars on the velocity dispersion than the
others. Part of the discrepancy could be due to
our estimate and the estimate by \citet{heckman85} being based on
different spectral regions. The blue part of the spectrum, used by
\citet{heckman85}, could be more contaminated by AGN continuum than
the red part, affecting velocity dispersion estimates.

\begin{figure}
\begin{center}
\includegraphics{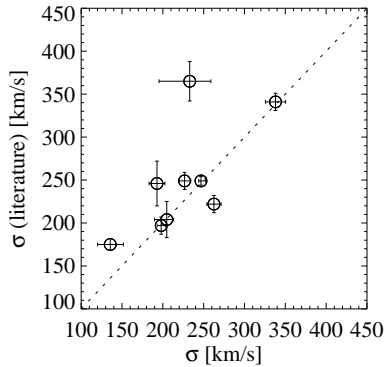}
\caption{Comparison with dispersions from the literature.}
\label{fig:lit_comparison}
\end{center}
\end{figure}


\section{Discussion}

\subsection{Conversion to AGN parameters}
\label{section:agnparam}

We convert the velocity dispersions derived from the spectral fitting
to a common physical scale following the scheme by
\citet{jorgensen95}, and thereafter to BH mass using the $M_{\rm
  BH}$--$\sigma_{*}$ relation from \cite{tremaine02}. We find that the
sample spans a relatively narrow range in BH mass, from
2.2$\times$10$^{7}$ to 1.1$\times$10$^{9}$ M$_{\odot}$.  Assuming that
we are able to recover the velocity dispersions to within an accuracy
of $\approx10$\% as shown by the simulations, and including a scatter
of 0.3 dex in $\log M_{\rm BH}$ (Tremaine et al.\ 2002) around the
$M_{\rm BH}-\sigma_{*}$ relation, we estimate that we are able to
determine BH masses to an accuracy of approximately 0.35 dex.

Narrow emission line fluxes for [\ion{O}{ii}] and [\ion{O}{iii}] were
determined from our 4\arcsec\, slit spectra, and some were also taken
from the literature, see columns 11 and 12 of Table~\ref{table:t1}.
Under the assumption that the narrow emission lines are photoionzied
by the UV continuum from the AGN, we convert to total luminosity in
the narrow-line region using the relation by \cite{rs91}, $L_{\rm NLR}
\approx 3\left(3L_{\rm [OII]}+1.5 L_{\rm [OIII]}\right)$.  In cases
where only one of the two line fluxes are known (for FRIIs), we apply
a relation between $L_{\rm [OII]}$ and $L_{\rm [OIII]}$ from
\citet{grimes04}, whereas for FRIs we take $L_{\rm [OII]} = L_{\rm
  [OIII]}$.  If there is an upper limit on the flux, we take the line
flux to be equal to the limit.  The narrow line flux is converted to
photoionizing luminosity, $Q_{\rm phot}$, by assuming a covering
factor of $\kappa=0.005$ \citep{willott99} for the narrow-line gas.
We estimate that the uncertainty in $\log Q_{\rm phot}$ is rougly $\pm
1$ dex, which includes uncertainties related to the covering factor
for the narrow-line gas and the fact that narrow lines may also be
powered by shocks \citep{ds95,inskip02}.

In order to estimate the average kinetic power transported by the jets
to the lobes, we use the relation of \cite{willott99}, $Q_{\rm jet} =
3\times10^{38} L_{\rm 151}^{6/7}$ Watts, where $L_{\rm 151}$ has units
of 10$^{28}$ W\,Hz$^{-1}$\,sr$^{-1}$. The radio luminosity at 151 MHz
for the sources in our sample was taken from Grime's
compilation\footnote{http://www-astro.physics.ox.ac.uk/\~\,sr/grimes.html}.
Given that Willott et al.'s relation applies to a ``typical'' radio
galaxy of median age, $\pm 1$ dex is a reasonable estimate for the
uncertainty in $\log Q_{\rm jet}$.

We also derive the dimensionless ratio between the jet luminosity and
the photoionizing luminosity, $\log (Q_{\rm jet}/Q_{\rm phot})$, and
estimate an uncertainty in this parameter of approximately $\pm 1.5$
dex.  The $Q_{\rm jet}/Q_{\rm phot}$ ratio can be seen as the ratio of
kinetic to radiative energy output by the AGN, and indicates how
effectively jets are formed. A small ratio implies that the energy
output is dominated by thermal radiation from an accretion disk,
whereas a large ratio indicates that most of the energy goes into
forming jets, i.e.\ to kinetic energy, and that the accretion process
thus may be radiatively inefficient.

\begin{figure*}
\centering
\includegraphics[width=17cm]{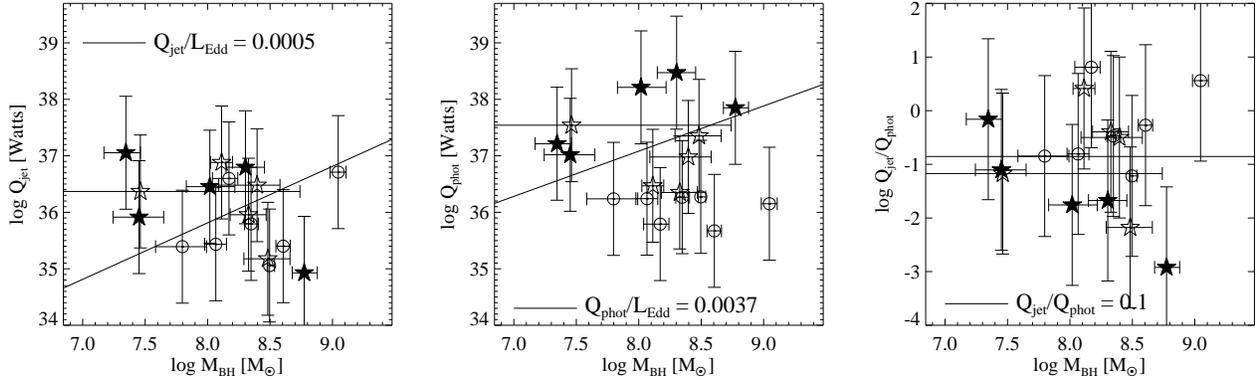}
\caption{From left to right, jet power, photoionizing luminosity and
  their ratio as a function of BH mass. An uncertainty of $\pm1$ dex
  in $Q_{\rm jet}$ and $Q_{\rm phot}$ has been assumed, and the 95 \%
  confidence interval on BH mass is used. Symbols are as in
  Fig.~\ref{figure:lo}.}
\label{figure:agnparam}
\end{figure*}


\subsection{The Ledlow-Owen relation}
\label{section:lorelation}

The parameters $Q_{\rm jet}$, $Q_{\rm phot}$ and their ratio are
plotted as a function of BH mass in Fig.~\ref{figure:agnparam}. The
line of separation between
FRIs and FRIIs, $L_{\rm radio} \propto L_{\rm optical}^{1.8}$ \citep{lo96},
can be written as
\begin{equation}
\log\left(\frac{P_{\rm 1.4}}{{\rm W\,Hz^{-1}}}\right) = -0.67 M_{R} + 10.13,
\label{eq:lo_eqn}
\end{equation}
\noindent
where $M_{R}$ is the optical absolute magnitude of the host and
$P_{\rm 1.4}$ is the radio power at 1.4 GHz. By using the relation by
\cite{md01} which relates $M_{R}$ to $M_{\rm BH}$ and the relation
from Willott et al.\ (1999) linking $P_{\rm 1.4}$ to $Q_{\rm jet}$, we
obtain
\begin{equation}
\left(\frac{Q_{\rm jet}}{{\rm W}}\right) = 6.17\times10^{27}\left(\frac{M_{\rm BH}}{M_{\odot}}\right)^{0.92}. 
\label{eq:qjet}
\end{equation}
This translates into $Q_{\rm jet} \simeq 5\times10^{-4} L_{\rm Edd}$,
implying that the FRI/FRII transition occurs at a ratio of
$\simeq$0.0005 between jet power and Eddington luminosity if $M_{\rm
  BH}\approx10^{8}$ M$_{\odot}$. \cite{gc01} derive a ratio of 0.015,
but this is probably erroneous.\footnote{The factor in eq.~2 of
  Ghisellini \& Celotti (2001) should be 3$\times$10$^{15}$ instead of
  3$\times$10$^{21}$.}  The left-hand panel of
Fig.~\ref{figure:agnparam} shows $Q_{\rm jet}$ as a function of
$M_{\rm BH}$, and the Ledlow-Owen relation in the form $Q_{\rm jet} =
5\times10^{-4} M_{\rm BH}$ is indicated with a solid line and seems to
fit the data well.

The Ledlow-Owen relation can also be written in terms of photoionizing
luminosity through the radio-optical correlation for radio galaxies
\citep{willott99}. Doing this, we find $Q_{\rm phot} \simeq 0.0037
L_{\rm Edd}$ for the typical BH mass in our sample. (\citet{gc01}
derive $Q_{\rm phot} \simeq 0.005 L_{\rm Edd}$). The solid line in the
middle panel of Fig.~\ref{figure:agnparam} shows the Ledlow-Owen
relation recast in the two variables $Q_{\rm phot}$ and $M_{\rm BH}$
and can be seen to fit the transition region quite well.
However, the separation of FRIs and FRIIs appears to be cleaner in the
diagram of $\log Q_{\rm phot}$ vs $\log M_{\rm BH}$, probably because
$Q_{\rm jet}$ is more influenced by external factors, such as
conditions in the interstellar medium of the host.  If the accretion
efficiency is approximately 10\%, and assumed to be in the form of a
standard optically thick, geometrically thin accretion disk
\citep{ss73}, the relation $Q_{\rm phot} \approx 0.004 L_{\rm Edd}$
indicates that the FRI/FRII transition occurs at $\sim 4$\% of the
Eddington rate.

The transition can also be described as occurring at an almost fixed
ratio between $Q_{\rm jet}$ and $Q_{\rm phot}$, i.e.\ at a fixed
``radio loudness''. Using the relations above, we find $Q_{\rm
  jet}/Q_{\rm phot} \approx 0.1$, shown as a solid line in the
rightmost panel of Fig.~\ref{figure:agnparam}, but the sources do not
separate as well in this diagram as in the other two.

Both AGN and X-ray binaries (XRBs) are associated with accretion onto
compact objects and the production of jets. The two have therefore
been compared at several occations
\citep[e.g.][]{mhd03,falcke04,maccarone03,mchardy06}.  XRBs are characterized by
three different spectral states \citep[see e.g.][]{gallo03}. The
low/hard state has a hard power-law spectrum, a weak thermal component
from an accretion disk/flow, and steady jet emission is seen at radio
wavelengths. FRIs have been suggested as the high BH mass analogue of
the low/hard state in XRBs \citep[see e.g.][]{mhd03,falcke04} and
having accretion flows with low accretion rates and inefficient
cooling, such as advection-dominated accretion flows
\citep{rees82,ny94}.  FRIIs, on the other hand, are thought of as
analogous to the high/soft and very high state XRBs with accretion
disks modeled as standard, optically thick, geometrically thin disks
\citep{ss73}. XRBs in the high/soft state show no strong or steady
radio emission, and jet formation is thought to be inhibited. In the
so-called very high state, thought to occur at high accretion rates,
XRBs show transient jet phenomena.

If, when we derive the relations in Eqs. \ref{eq:lo_eqn} and
\ref{eq:qjet}, keep the BH mass dependence instead of substituting the
mean black hole mass of the sample, we find that $Q_{\rm phot}/L_{\rm
  Edd} \propto (M_{\rm BH}/M_{\odot})^{-0.14}$ and $Q_{\rm jet}/L_{\rm
  Edd} \propto (M_{\rm BH}/M_{\odot})^{-0.08}$.  Hence both the
photoionizing and jet luminosity in Eddington units at the FR/FRII
transition is weakly dependent on BH mass.  It is too weak to be
noticeable in a sample of radio galaxies, but when extrapolated to
XRBs there should be roughly a factor of 10 difference.  From the
Ledlow-Owen relation we therefore expect that the FRI/FRII transition
luminosity in Eddington units is about a factor of 10 lower than for
spectral state transition in XRBs.  This does indeed seem to be
observed. For the FRI/FRII transition we observe transition
luminosities of $\approx 0.4$\% of the Eddington luminosity, see
Fig.~\ref{figure:agnparam}.  The transition luminosity of XRBs from
the high/soft to the low/hard state is $\approx2$\% of the Eddington
luminosity (may be up to four times higher for the low/hard to
high/soft transition) \citep{mgf03}.  The concurrence of observed
transition luminosities with those predicted from radio galaxy scaling
relations may indicate that the transition between low/hard and
high/soft states in XRBs is indeed similar in nature to the FRI/FRII
transition. \citet{maccarone03} discusses that neutron star XRBs may
have higher transition luminosities than BH XRBs, also indicating that
the transition luminosity may depend on the mass of the compact
object.

If FRIIs are analogues to the very high state XRBs with thermally
unstable and radiation pressure dominated accretion disks
\citep{meier01}, the FRI/FRII transition could correspond to a switch
between the low/hard and the very high state, without going through
the high/soft state. \citet{meier01} offers a solution which has the
transition luminosity relative to the Eddington luminosity
proportional to $M_{\rm BH}^{-1/8}$ (see also
\citealt{chen95,merloni03}). This may suggest that the FRI/FRII
transition is due to a transition similar to that between the low/hard
state and the very high state in XRBs.  There is some inconsistency in
the implied transition luminosities in this scenario because the very
high state in XRBs is thought to occur at a (bolometric) luminosity of
20--30\% of the Eddington luminosity \citep{mgf03,meier01}, whereas we
find $<1$\% for the radio galaxies.  The discrepancy could be due to
our different methods of estimating the bolometric/photoionizing
luminosity, but another possibility is that the physical conditions
(e.g.\ density) in the accretion flow close to the BH are different
for radio galaxies and XRBs.

Sources that do not follow the Ledlow-Owen relation may be the ones in
clusters or high density environments \citep{marchesini04}.  Besides
external density in this case probably being a dominant factor for the
radio morphology, another possible explanation is that mergers or
collisions in cluster galaxies may have altered the spin of the BH
\citep{np98} and BH spin is thought to be important for the formation
of jets.

The Ledlow-Owen relation nevertheless seems to be an indication that
the FRI/FRII morphology is largely determined by nuclear AGN
parameters. In hindsight it is perhaps not surprising that the
empirical scaling relations used by \citet{bicknell95} to explain the
Ledlow-Owen relation relate to the central stellar velocity
dispersion.


\section{Summary and conclusions}

We have investigated BH masses and narrow emission line luminosities
in a sample of FRI and FRII radio galaxies.  BH masses were estimated
via the $M_{\rm BH}$--$\sigma$ relation by determining central stellar
velocity dispersions from medium resolution spectra, and photoionizing
luminosities were estimated from spectrophotometry of the narrow
[\ion{O}{II}]3727 {\AA} and [\ion{O}{III}]5007 {\AA} emission lines.
The measured velocity dispersions span the range 130--340 km\,s$^{-1}$
with a mean of $\approx 215$ km\,s$^{-1}$. Including the scatter in
the $M_{\rm BH}$-$\sigma$ relation, we estimate that we are able to
determine BH masses with an accuracy of $\approx 0.35$ dex.  We find
FRIs and FRIIs to have the similar BH mass distributions, confirming
previous results \citep[e.g.][]{mclure04,marchesini04}.

The data fit the Ledlow-Owen relation well when it is rewritten in
terms of jet luminosity, photoionizing luminosity and BH mass
\citep{gc01}.  The separation of FRIs and FRIIs appears cleaner in the
$Q_{\rm phot}$--$M_{\rm BH}$ plane than in the $Q_{\rm jet}$--$M_{\rm
  BH}$ plane, indicating that FRIs and FRIIs are more easily separated
in terms of nuclear parameters than in terms of parameters such as
$Q_{\rm jet}$ which may be more influenced by external factors.  However, it is
still possible that the FR/FRII transition may be caused by a mixture of both 
nuclear and external factors.

We discuss whether the FRI/FRII transition can be compared to spectral
state transitions in XRBs, and highlight the fact that the Ledlow-Owen
relation predicts a BH mass dependence in the transition luminosity in
Eddington units. This implies that radio galaxies change from FRI to
FRII at luminosities (in Eddington units) that are a factor of 10
lower that those of XRBs.

\begin{acknowledgements}
  The authours thank the referee for comments which helped to improve
  the original manuscript. This research has made use of the NASA/IPAC
  Extragalactic Database (NED) which is operated by the Jet Propulsion
  Laboratory, California Institute of Technology, under contract with
  the National Aeronautics and Space Administration. The authors would
  also like to thank the maintainers of the DRAGN web page ({\em
    http://www.jb.man.ac.uk/altas}) which was used in this work.
\end{acknowledgements}

\bibliographystyle{aa}


\end{document}